\documentclass[onecolumn,showpacs,amsmath,amssymb,pre]{revtex4-1}

\usepackage{graphicx}
\usepackage{color}
\usepackage{dcolumn}
\usepackage{amsmath}
\usepackage{subfigure}

\usepackage{sidecap}
\usepackage{dsfont}

\usepackage{mathrsfs}
\usepackage{amsfonts}
\usepackage{indentfirst}
\usepackage{bm}

\usepackage[colorlinks,citecolor=blue]{hyperref}

\newcommand{\be}{\begin{equation}}
\newcommand{\ee}{\end{equation}}
\newcommand{\bey}{\begin{eqnarray}}
\newcommand{\eey}{\end{eqnarray}}
\newcommand{\bw}{\begin{widetext}}
\newcommand{\ew}{\end{widetext}}

\newcommand{\ra}{\rangle}
\newcommand{\la}{\langle}

\newcommand{\ba}{\begin{array}}
\newcommand{\ea}{\end{array}}
\newcommand{\bi}{\begin{itemize}}
\newcommand{\ei}{\end{itemize}}
\newcommand{\bem}{\begin{enumerate}}
\newcommand{\eem}{\end{enumerate}}

\begin{document}

\title{Multifractality and excited-state quantum phase transition in ferromagentic
spin-$1$ Bose-Einstein condensates}

\author{Zhen-Xia Niu$^{1}$}\email{niuzhx@zjnu.edu.cn}
\author{Qian Wang$^{1,2}$}\email{qwang@zjnu.edu.cn}

\affiliation{$^{1}$Department of Physics, Zhejiang Normal University, Jinhua 321004, China \\
$^2$CAMTP-Center for Applied Mathematics and Theoretical Physics, University of Maribor,
Mladinska 3, SI-2000, Maribor, Slovenia}

\date{\today}

\begin{abstract}

Multifractality of quantum states plays an important role 
for understanding numerous complex phenomena observed in different branches of physics.
The multifractal properties of the eigenstates allow for charactering various phase transitions.   
In this work, we perform a thoroughly analysis of the impacts of an 
excited-state quantum phase transition (ESQPT) 
on the fractal behavior of both static and dynamical wavefunctions 
in a ferromagentic spin-$1$ Bose-Einstein condensate (BEC).
By studying the features of the fractal dimensions, we show how
the multifractality of eigenstates and time evolved state are affected by the presence of ESQPT.
Specifically, the underlying ESQPT leads to a strong localization effect, which in turn enables us
to use it as an indicator of ESQPT.
We verify the ability of the fractal dimensions to probe the occurrence of ESQPT
through a detailed scaling analysis. 
We also discuss how the ESQPT manifests itself in the fractal 
dimensions of the long-time averaged state. 
Our findings further confirm that the multifractal analysis is a powerful tool 
for studying of phase transitions in quantum many-body systems and
also hint an potential application of ESQPTs in burgeoning 
field of state preparation engineering.

\end{abstract}

\maketitle

\section{introduction}

Analyzing the multifractality of quantum states is central to understanding several fundamental
questions in various areas of physics, ranging from quantum chaos
\cite{Berry1977,Shapiro1984,Kus1988,Izrailev1990,Backer2002,
Backer2007,Beugeling2018,Backer2019,Pausch2021a,Tomasi2020} 
and quantum statistics \cite{Deutsch1991,Srednicki1994,Alessio2016,Borgonovi2016,Garrison2018} 
to condensed matter physics \cite{Mirlin2000b,Evers2008,Nandkishore2015,Fabien2018}.
Apart from extremely simple cases, quantum states in any computation basis are usually 
have a larger number of expansion coefficients.
Consequently, the multifractality of quantum states is naturally encoded in 
the statistics of the expansion coefficients.
The statistical properties of quantum states have been investigated in multiple ways. 
Among them, the inverse participation ratio is commonly employed to
characterize the quantum state statistics for both single-particle 
\cite{Evers2008, Wegner1980,Kramer1993,Evers2000,Calixto2015} 
and many-body systems
\cite{Rigol2010,Beugeling2015,Bera2015,Herrera2015,
Misguich2016,Tsukerman2017,Lezama2021,Frey2024}.
The quantum state statistics can also be explored via the Shannon and R\'{e}nyi entropies 
\cite{Stephan2009,Stephan2010,Santos2010a,Santos2010b,Stephan2011,Luitz2014,Misguich2017},
which are a generalization of the inverse participation ratio.
More characterizations of the quantum state statistics are revealed by the so-called
multifractal analysis, which focuses on the studying of fractal dimensions \cite{Hentschel1983,Halsey1986}
and is the topic of the present work.
It was known that the multifractal analysis is 
a powerful tool for understanding of the Anderson localization 
\cite{Mirlin2000a,Evers2008,Rodriguez2011}
and the properties of eigenstates in quantum systems
\cite{Pausch2021a,Beugeling2018,Backer2019,Serbyn2017,Mace2019,
Luitz2020,Pausch2021b,Pausch2022,Atas2012}.  
In particular, the multifractal analysis has also be used to
uncover signatures of different phase transitions exhibited by quantum many-body systems 
\cite{Lindinger2019,Sierant2022}.

The study of excited-state quantum phase transitions (ESQPTs) 
\cite{Caprio2008,Stransky2014,Cejnar2021} 
has recently attracted a lot of interest in both theoretical 
\cite{Brandes2013,Magnani2014,Bastidas2014,Puebla2016,Fernandez2017,
WangQW2020,Corps2021,Feldmann2021,Stransky2021,Khalouf2022,HuG2023} 
and experimental \cite{Dietz2013,Meyer2023} works.
ESQPTs are a generalization of the ground state quantum phase transition and 
are signified by the nonanalytical behaviors in the density of states.
The affects of ESQPTs have been extensively investigated in various many-body systems,
such as the Lipkin-Meshkvo-Glick (LMG) model
\cite{Relano2008,PerezF2009,WangQ2019,WangQP2019,YuanZG2015,
WangQP2021,Nader2021,WangQP2021b,Santos2017,WangH2017,Mzaouali2021,
Santos2015,Santos2016,Gamito2022,Engelhardt2015,Gamito2023} 
the Dicke and Tavis-Cumming models 
\cite{Engelhardt2015,PerezF2011,Puebla2013,Lobez2016,Kloc2021,Corps2023}, 
the Kerr nonlinear oscillator \cite{WangQW2020,Chavez2023}, 
the interacting boson models \cite{Macek2019,Dong2021},
and spinor Bose-Einstein condensates (BECs)
\cite{Feldmann2021,Cabedo2021,ZhouL2023,ZhenXW2023}.
It was also found that ESQPTs have a connection to the onset of quantum chaos
\cite{PerezFP2011,Garca2021} and can be employed to engineer cat state \cite{CorpsA2022}.  
A detailed review of features and applications of ESQPTs are given in Ref.~\cite{Cejnar2021}.

It is natural to ask what is the impacts of ESQPTs on the multifractality of quantum states.
Several previous studies have been explored this question by means of 
the participation ratio \cite{Santos2015,Santos2016,Gamito2022,Jamil2022}.
They have shown that the presence of ESQPT can strongly change the structure of the eigenstates,
resulting in a highly localized state around the critical point of ESQPT.
However, a fully understanding of how ESQPTs affect the fractal dimensions of eigenstates, 
and particularly the time evolved state remains less known. 
The present work is an extension of previous works with
aim to provide a thoroughly examination of the impacts of ESQPT 
on the fractal properties of quantum states from both static and dynamic aspects.

In this work, we address the following questions: "How the fractal dimensions of the eigenstates and
time evolved state vary as the system passes through the critical point of ESQPT? 
And, whether the behaviors of the fractal dimensions can be used to 
identify the occurrence of ESQPT?"
To this end, we perform a detailed multifractal analysis in a ferromagnetic spin-$1$ BEC.
The spin-$1$ BEC is a highly tunable system and exhibits rich phases 
for both ground and excited states 
\cite{LiuY2009,Kawaguchi2012,Stamper2013,Meyer2023,
Dag2018,Feldmann2021,Bookjans2011,Jacob2012}.
It has been widely used to investigate the static and dynamical properties 
of different phase transitions 
\cite{ZhangNJP2003,Sadler2006,Lamacraft2007,Damski2007,
Anquez2016,Vinit2017,XueM2018,Feldmann2018,YangH2019,TianT2020}. 
In particular, the recent theoretical and experimental studies have been verified the existence of 
ESQPTs in the spin-$1$ BEC \cite{Feldmann2021,Meyer2023}. 
This makes it a suitable system for our purpose. 

We show that the fractal dimensions of the eigenstates exhibit 
a sudden dip near the ESQPT critical energy.
This in turn indicates that the fractal dimensions behaves as the probe of ESQPT.  
Further scaling analysis demonstrates that the eigenstate with energy close to the ESQPT critical energy
becomes a localized state in the thermodynamic limit.
To uncover the effects of ESQPT on the multifractality of time evolved state, 
we consider a sudden quench process.
We show that the underlying ESQPT leaves strong imprint in the fractal dimensions of time evolved state.
As a result, both short- and long-time dependences of the fractal dimensions undergo a remarkable
change when we straddled the critical point of ESQPT. 
We also discuss how ESQPT affects and gets reflected in the fractal properties
of the long-time averaged state.

The rest of this article is organized as follows.
In Sec.~\ref{Second}, we introduce the fractal dimensions 
and provide a brief review of some basic features of the ferromagnetic spin-$1$ BEC.
In Sec.~\ref{Third}, we present our main results.
Specifically, Sec.~\ref{ThirdA} is devoted to discussing the behaviors of the 
fractal dimensions of the eigenstates.
Thereafter, we carry out the multifractal analysis for both time evolved 
and long-time averaged states in Sec.~\ref{ThirdB}.
We finally summarize our studies in Sec.~\ref{Fourth} with several remarks.

 \begin{figure}
  \includegraphics[width=\columnwidth]{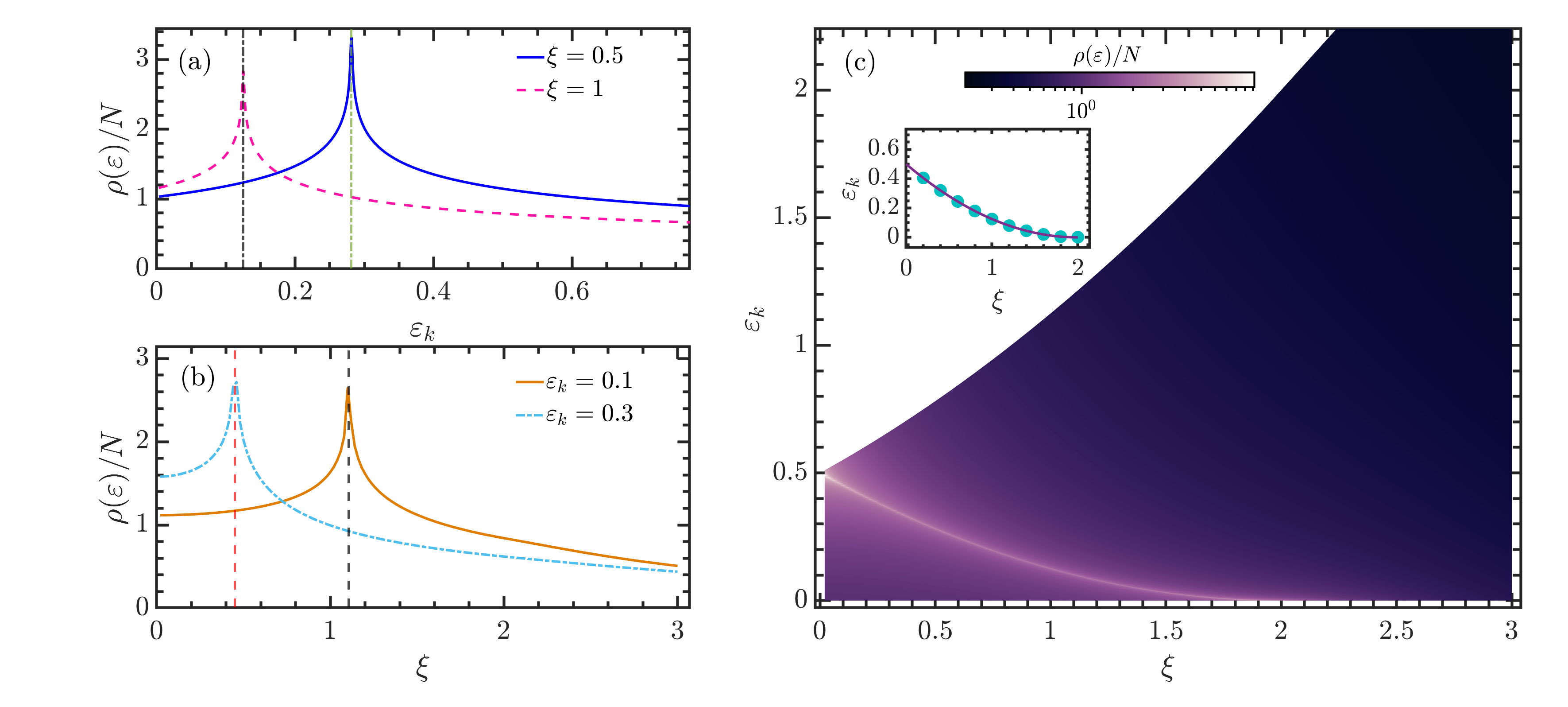}
  \caption{(a) Rescaled DOS, $\rho(\varepsilon)/N$, 
  as a function of normalized eigenenergy, $\varepsilon_k=(E_k-E_0)/N$, for different values of $\xi$.
  Here, $E_k$ is the energy of the $k$ eigenstate with $E_0$ being the energy of the ground state.
  The gray and green vertical dot-dashed lines denote the 
  critical energies obtained from Eq.~(\ref{CriticalEg}).
  (b) $\rho(\varepsilon)/N$ as a function of $\xi$ for different energies. 
  The red and gray vertical dashed lines
  mark the critical value of $\xi$ for the corresponded energies (cf.~Eq.~(\ref{CriticalEg})].
  (c) $\rho(\varepsilon)/N$ as a function of $\xi$ and $\varepsilon_k$.
  Inset: Position of local peaks in the DOS as a function of  $\xi$ and $\varepsilon_k$.
  The purple curve plots the analytical result given by Eq.~(\ref{CriticalEg}).
  Other parameter: $N=5000$.}
  \label{DoSFigs}
 \end{figure}

\section{Premilinaries: Fractal dimensions and Model}\label{Second}

\subsection{Fractal dimensions}\label{FrDs}

Fractal dimensions characterize the multifractality of the quantum state and play a crucial role for understanding
various complex phenomena observed in many-body systems, 
including Anderson and many-body localization transitions 
\cite{Mirlin2000a,Mirlin2000b,Evers2008,Rodriguez2011,Serbyn2017,Mace2019,Luitz2020},  
different quantum phase transitions \cite{Atas2012,Luitz2014,Misguich2017,Sierant2022,Lindinger2019}, 
and quantum chaos \cite{Backer2019,Pausch2021a,Tomasi2020,WangQ2021,Pausch2021b,Pausch2022}.
To define the fractal dimensions, let us first expand a quantum state $|\phi\ra$ in an orthonormal
basis $\{|\alpha\ra\}$ with size $\mathcal{N}$ as
$|\phi\ra=\sum_{\alpha=1}^\mathcal{N} c_\alpha|\alpha\ra$,
where $c_\alpha=\la\alpha|\phi\ra$ and satisfies $\sum_\alpha|c_\alpha|^2=1$.
Then, the finite-size fractal dimensions of $|\phi\ra$ is defined by
\cite{Lindinger2019,Backer2019,Pausch2021a}
\be\label{Frdm}
  D_q=\frac{S_q}{\ln\mathcal{N}}, 
\ee
where $S_q=\ln\left(\sum_\alpha|c_\alpha|^{2q}\right)/(1-q)$ is the $q$-dependent R\'{e}nyi
(or participation) entropy and $q\in\mathbb{R}^+$.
It is known that $D_q$ is varied in an interval $D_q\in[0,1]$ and decreases with increasing $q$.
Finally, the values of $D_q$ in the limit $\mathcal{N}\to\infty$ provide the fractal dimensions, 
$\widetilde{D}_q=\lim_{\mathcal{N}\to\infty}D_q$ \cite{Mirlin2000a,Backer2019}.
 
For a fully delocalized state, we have $|c_\alpha|^2=1/\mathcal{N}$ as $\mathcal{N}\to\infty$ and thus $D_q^\infty=1$. 
On the contrary, the extremely localized state leads to 
$c_{\alpha'}=1$ and $c_\alpha=0$ for $\alpha\neq\alpha'$, resulting in $D_q^\infty=0$.
The states with $0<D_q^\infty<1$ are the so-called mutlifractal states, 
which are extended but nonergodic in the given basis.

In this work, we focus on $D_q$ with $q=1,2$, and $\infty$.
As $S_1$ turns into the Shannon entropy $S_h=-\sum_\alpha|c_\alpha|^2\ln|c_\alpha|^2$ for $q=1$,
the dimension $D_1$ quantifies the scaling of $S_h$ and is called as the information dimension. 
At $q=2$, we have $S_2=\ln(\sum_\alpha|c_\alpha|^4)^{-1}$, which is the logarithm of the
participation ratio. 
Hence, $D_2$ characterizes how the participation ratio varies with $\mathcal{N}$.
Finally, due to $S_\infty=\lim_{p\to\infty}S_q=-\ln p_{max}$ with 
$p_{max}=\max_\alpha|c_\alpha|^2$, we thus have $D_\infty=-\ln p_{max}/\ln\mathcal{N}$.

 \begin{figure}
  \includegraphics[width=\columnwidth]{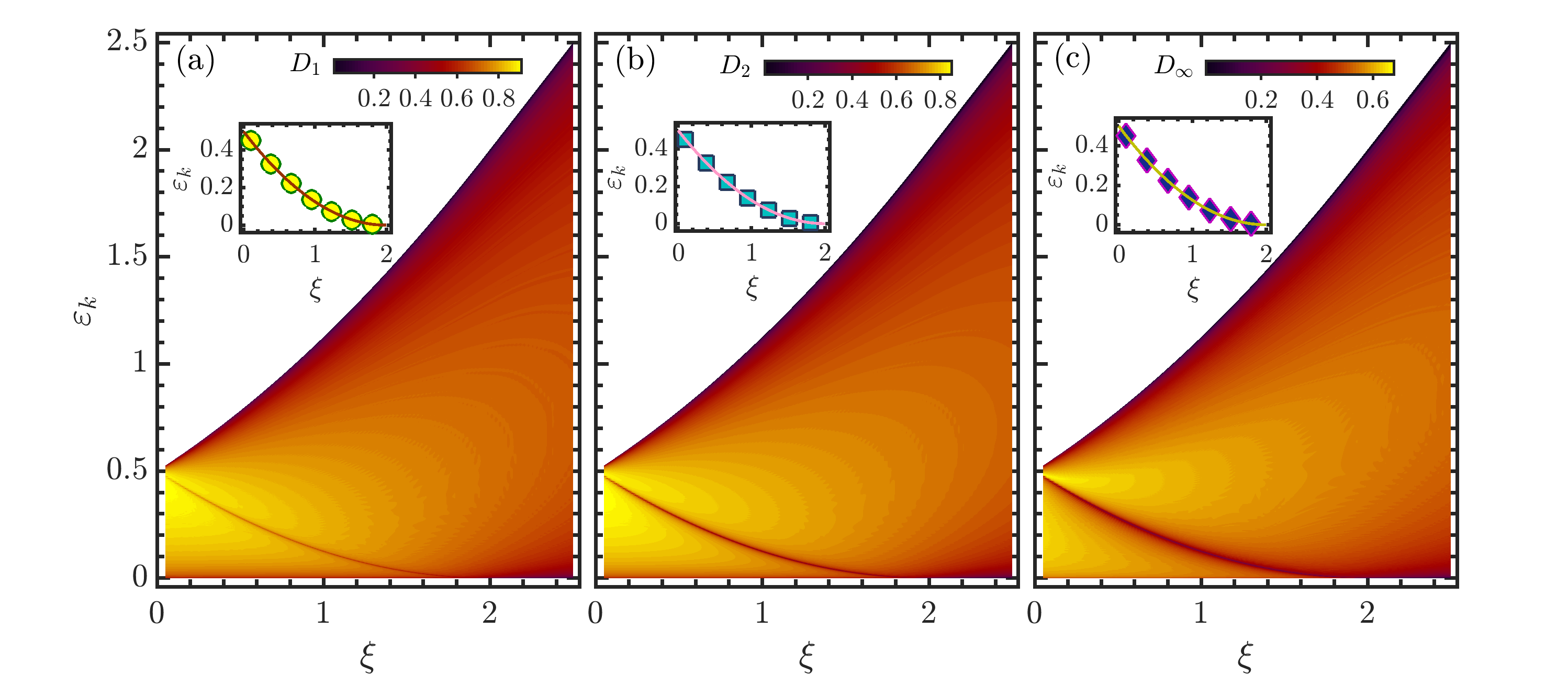}
  \caption{(a)-(c): Finite-size fractal dimensions $D^{(k)}_1$ (a), $D^{(k)}_2$ (b), 
  and $D^{(k)}_\infty$ (c) as a function of $\varepsilon_k$ and $\xi$.
  The inset in each panel plots how the position of dip in corresponding fractal dimension
  varies as a function of $\xi$ and $\varepsilon_k$.
  The solid curves denote the analytical result in Eq.~(\ref{CriticalEg}).
  Other parameter: $N=1000$.}
  \label{FDEks}
 \end{figure}

\subsection{Ferromagnetic spin-$1$ BEC}

 We consider a sufficiently weak trapped 
 spinor BEC of $N$ atoms with three spin states $m=0,\pm1$.
 This implies the validity of the single-mode approximation (SMA) 
 \cite{Kawaguchi2012,Stamper2013}, 
 which assumes the same spatial mode for all spin states. 
 Consequently, the Hamiltonian of spin-$1$ BEC under the SMA reads 
 \cite{Feldmann2018,XueM2018,Feldmann2021}
 \be \label{SpinH}
  \frac{H}{|c|}=\frac{\mathrm{sign}(c)}{N}\left[N_0(N_1+N_{-1})+(a_0^2a_1^\dag a_{-1}^\dag
  +a_{-1}a_1a_0^{\dag2})+\frac{1}{2}(N_1-N_{-1})^2\right]+\xi(N_1+N_{-1}),
 \ee
 where $a_m^\dag$ ($a_m$) is the bosonic creation (annihilation) operator for spin states,
 $N_m=\sum_ma_m^\dag a_m$ denotes the number operator of the $m$th state,
 and $c$ sets the energy scale of the spin-dependent interaction with $c<0\ (c>0)$ corresponding to
 ferromagnetic (antiferromagnetic) spin-dependent interactions.
 Here, $\xi=q/|c|$ is the rescaled quadratic Zeeman shift, 
 which can be tuned via the microwave dressing. 
 
 The Hamiltonian (\ref{SpinH}) conserves 
 the total magnetization $M=N_{1}-N_{-1}$ and the parity $\Pi=(-1)^{N_0}$ \cite{Feldmann2021,ZhenXW2023}.
 This allows us to restrict our study in the subspace with $M=0$ and $\Pi=1$.
 As a result, the dimension of the Hilbert space is $\mathcal{N}=N/2+1$ for even $N$.
 Moreover, we further focus on the case of ferromagnetic interaction and $\xi\geq0$. 
 
 It is known that the ferromagnetic spin-$1$ BEC undergoes a quantum phase transition (QPT) 
 as $\xi$ passes through the critical point $\xi_c=2$, which separates 
 the broken-axisymmetry phase ($0<\xi<2$) from the
 polar phase ($\xi>2$) \cite{Kawaguchi2012,Stamper2013}. 
 There are numerous works are devoted to exploring both the static 
 and dynamical properties of the QPT
 in the spin-$1$ BECs 
 \cite{XueM2018,ZhangNJP2003,Sadler2006,Lamacraft2007,Damski2007,
 Bookjans2011,Jacob2012,Anquez2016}. 
 Apart from the ground state QPT, the Hamiltonian (\ref{SpinH}) 
 also exhibits an ESQPT \cite{Feldmann2021,ZhenXW2023}, 
 which is characterized by the singularities in the density
 of the states (DOS) \cite{Cejnar2021}. 
 
 In Fig.~\ref{DoSFigs}(a), we show how the rescaled DOS, $\rho(\varepsilon)/N=\sum_k\delta(\varepsilon-\varepsilon_k)$,
 varies as a function of the normalized energy, $\varepsilon_k=(E_k-E_0)/N$, for several values of $\xi$. 
 Here, $E_k$ denotes the energy of the $k$th eigenstate with $k=0$ corresponding to the ground state.
 One can see that, regardless of the value of $\xi$, 
 DOS exhibits an obvious sharp peak at an excited energy, 
 which is dubbed as the critical energy of an ESQPT.
 The critical energy, denoted by $\varepsilon_c$, of spin-$1$ BEC depends on the value of $\xi$, 
 as illustrated in Fig.~\ref{DoSFigs}(a). 
 By utilizing the semiclassical (or mean-field) approach, one can find that 
 the explicit form of $\varepsilon_c$ is given by \cite{ZhenXW2023}
 \be \label{CriticalEg}
   \varepsilon_c=\frac{(2-\xi)^2}{8},
 \ee
 with $0\leq\xi\leq2$.
 The ESQPT in ferromagnetic spin-$1$ BEC can also be revealed by  
 the behavior of $\rho(\varepsilon)/N$ as a function of $\xi$ with fixed energy. 
 In Fig.~\ref{DoSFigs}(b), we plot the variation of $\rho(\varepsilon)/N$ 
 with $\xi$ for two different energies.
 It is obvious that the presence of ESQPT is signified by a sharp peak in DOS at certain value of $\xi$.
 For a given energy, the value of $\xi$, at which the DOS presents 
 local maximum, can be obtained from Eq.~(\ref{CriticalEg}). 
 We further plot the overall behavior of $\rho(\varepsilon)/N$ 
 as a function of $\xi$ and $\varepsilon_n$ in Fig.~\ref{DoSFigs}(c). 
 The existence of ESQPT for $0<\xi<2$ is clearly marked by the local maxima in the DOS.
 Moreover, the position of local peaks in the DOS as a function 
 of $\xi$ and $\varepsilon_n$ shows an excellent agreement
 with Eq.~(\ref{CriticalEg}), as seen in the inset of Fig.~\ref{DoSFigs}(c).  
 
 In the rest of this article, we perform a detailed exploration 
 how the ESQPT affects the multifractality of the
 eigenstates and particularly the time evolved state by means of the multifractal analysis
 in a ferromagnetic spin-$1$ BEC. 
 Additionally, we also discuss how to identify the signatures of ESQPT in 
 both static and dynamical behaviors of the fractal dimensions.

 \begin{figure}
  \includegraphics[width=\columnwidth]{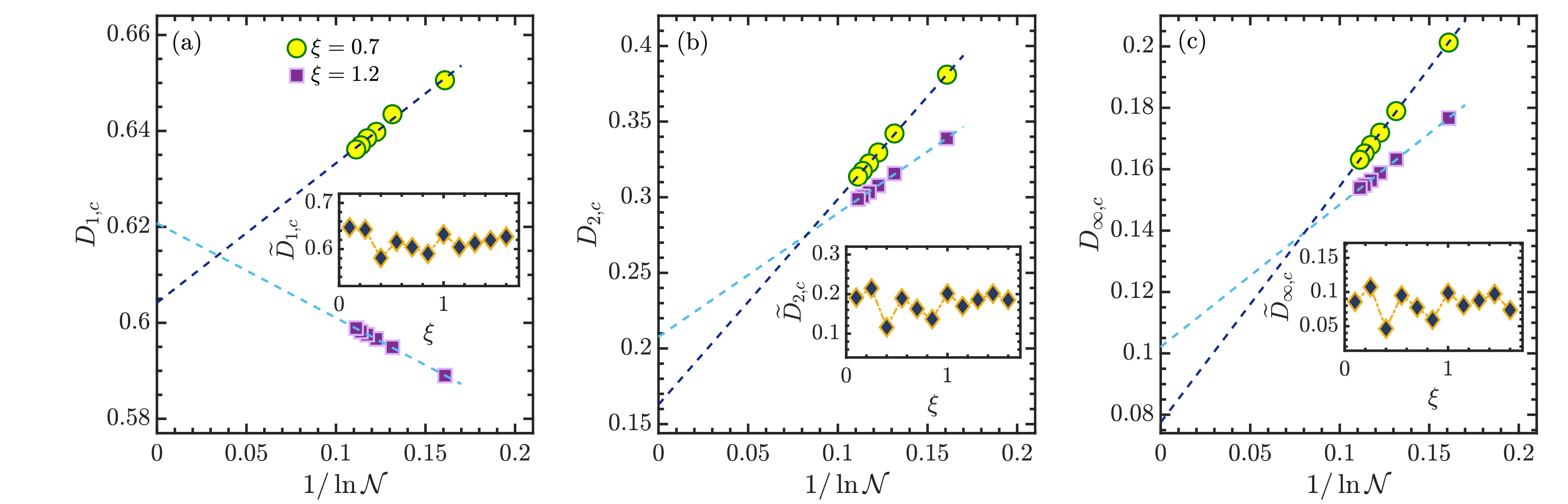}
  \caption{(a)-(c): Finite-size fractal dimensions at the critical energy, 
  $D_{q,c}=D_q(\varepsilon_k\simeq\varepsilon_c)$ 
  with $q=1$(a), $q=2$ (b), and $q=\infty$ (c) as a function of $1/\ln\mathcal{N}$
  for several $\xi$ values [see the legend in panel (a)].
  The dashed lines in each panel represent the linear fitting of the form
  $D_{q,c}=f_q/\ln\mathcal{N}+\widetilde{D}_{q,c}$ with $f_q$ and $\widetilde{D}_{q,c}$
  depending on the value of $\xi$.
 The inset in each panel plots how the fractal dimension 
 $\widetilde{D}_{q,c}=\lim_{\mathcal{N}\to\infty}D_{q,c}$ varies as a function of $\xi$.
  }
  \label{SFDs}
 \end{figure}

\section{ESQPT and fractal dimensions}\label{Third}

In this section, we delve into the interplay between ESQPT and the fractal dimensions in 
the ferromagnetic spin-$1$ BEC from both static and dynamic aspects.
To analyze the fractal dimensions, it is necessary to choose a 
basis in which we expand our considered quantum state.
The choice of the basis is usually motivated by the physical problems.
For the spin-$1$ BEC with $N_1=N_{-1}$, 
it is natural to select the Fock basis $\{|N,n_0\ra\}$ with $|N,n_0\ra$ being the eigenstates 
of $N_0$, so that $N_0|N,n_0\ra=n_0|N,n_0\ra$.
As the Fock states coincide with the eigenstates of the system in the limit $\xi\to\infty$, 
on can expect that $D_q\simeq0$ for sufficient large value of $\xi$.

\subsection{Statics}\label{ThirdA}

Let us first investigate the multifractality of the $k$th 
eigenstate $|\varepsilon_k\ra$ of $H$ (\ref{SpinH}).
In the Fock basis, $|\varepsilon_k\ra$ can be expanded as
\be
  |\varepsilon_k\ra=\sum_{n_0=0}^{N}c_{n_0}^{(k)}|N,n_0\ra,
\ee
where $c_{n_0}^{(k)}=\la N,n_0|\varepsilon_k\ra$ and $\sum_{n_0}|c_{n_0}^{(k)}|^2=1$.
The finite-size fractal dimension of $|\varepsilon_k\ra$ is therefore given by
\be \label{FDEk}
   D_q^{(k)}=\frac{S_q^{(k)}}{\ln\mathcal{N}},
\ee
with $S_q^{(k)}=\ln\left(\sum_{n_0}|c_{n_0}^{(k)}|^{2q}\right)/(1-q)$
being the $q$-dependent R\'{e}nyi entropy of the $k$th eigenstate.

It was known that the expansion coefficients $|c_{n_0}^{(k)}|^2$ exhibit a particular property
at the critical energy of an ESQPT \cite{Santos2015,Santos2016,Santos2017,Gamito2022}. 
Hence, according to Eq.~(\ref{FDEk}), we can expect that 
the finite-size fractal dimensions should undergo a remarkable
change as the system passes through the critical energy.
This is indeed what we see in Figs.~\ref{FDEks}(a)-\ref{FDEks}(c), where 
$D_1, D_2$ and $D_\infty$ are respectively plotted as a function of $\xi$ and $\varepsilon_k$.
A sudden dip along the critical energies, regardless of the value of $q$, 
in the behaviors of $D_q$ is clear visible. 
Further verification is provided by the insets of Figs.~\ref{FDEks}(a)-\ref{FDEks}(c), where 
we show how the position of the dip in $D_q$ varies as a function of $\xi$ and $\varepsilon_k$.
A good agreement between the numerical and analytical results indicates that 
the presence of ESQPT leaves a strong imprint in the behavior of fractal dimensions.
Moreover, the sudden dip in fractal dimensions also implies that it can be used to detect ESQPTs.

 \begin{figure}
  \includegraphics[width=\columnwidth]{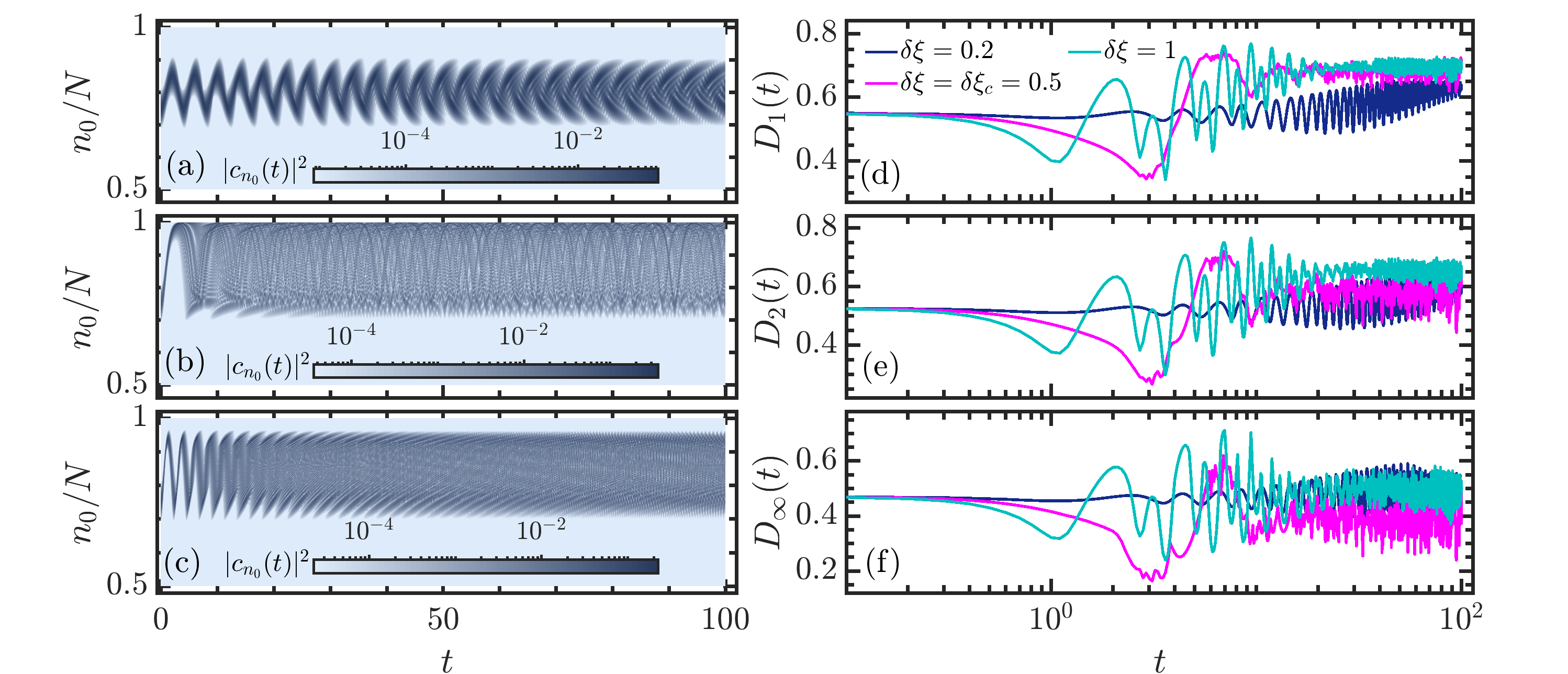}
  \caption{(a)-(c): Time evolution of the expansion coefficients $|c_{n_0}(t)|^2$ 
  for the time evolved state with $\delta\xi=0.2$ (a), $\delta\xi=0.5$ (b), and $\delta\xi=1$.
  (d)-(f): Time evolution of the fractal dimensions $D_q(t)$ in Eq.~(\ref{FrDt})
  corresponding to $q=1$ in (a), $q=2$ in (b), and $q=\infty$ in (c) for different quench strengths
  (see the legend in panel (a)). 
  Other parameters: $N=1000$ and $\xi_i=1$ with $\delta\xi_c=0.5$, obtained from 
  Eq.~(\ref{Cquench}).
  }
  \label{CfDqs}
 \end{figure}

To further reveal the impacts of ESQPT on the fractal dimensions, 
we analyze the scaling of $D_q$ in the limit $\mathcal{N}\to\infty$.
Figure \ref{SFDs} plots how the finite-size fractal dimensions at the critical energy,
denoted by $D_{q,c}$, evolves as a function of $1/\ln\mathcal{N}$
for different values of $\xi$. 
We see that the behaviors of $D_{q,c}$ are well captured by a linear scaling of the form
$D_{q,c}=f_q/\ln\mathcal{N}+\widetilde{D}_{q,c}$, 
with $\xi$-dependent $f_q$ and $\widetilde{D}_{q,c}$, regardless of the value of $q$. 
An explicit dependence of fractal dimensions $\widetilde{D}_{q,c}$ on $\xi$ 
are shown in the insets of Figs.~\ref{SFDs}(a)-\ref{SFDs}(c).
As $0<\widetilde{D}_{q,c}<1$ and it decreases with increasing $q$,
 the state at the ESQPT critical energy is the multifractal state.
This provides an application in the interdisciplinary field of preparing localized quantum states.

\subsection{Dynamics}\label{ThirdB}

 We continue our studies by focusing on 
 how the ESQPT affect the dynamics of $D_q$ and identify its dynamical signatures.
 To this end, we consider a sudden quench process.
 The system is initially prepared in the ground state, $|\varepsilon_0\ra$, of the Hamiltonian (\ref{SpinH}) with $\xi=\xi_i$. 
 At $t=0$, the value of $\xi$ is suddenly changed from its initial value $\xi_i$ 
 to a final value $\xi_f=\xi_i+\delta\xi$
 and the system starts to evolve according to the final Hamiltonian $H_f$.

For sudden quench process, one can vary the energy of the post-quenched 
system by tuning the value of $\delta\xi$.
Hence, there exists a particular quench, known as the critical quench and denoted by $\delta\xi_c$, 
which takes the post-quenched system to the critical energy of ESQPT.
The critical quench of our studied system can be obtained 
through the semiclassical (or mean-field) approach and the result is \cite{ZhenXW2023}
\be\label{Cquench}
  \delta\xi_c=1-\frac{\xi_i}{2},
\ee
with $0<\xi_i<2$.

At time $t$, the state of the system reads $|\psi(t)\ra=e^{-iH_ft}|\varepsilon_0\ra$.
Its expansion in the Fock basis can be written as 
$|\psi(t)\ra=\sum_{n_0}c_{n_0}(t)|N,n_0\ra$, where 
\be
  c_{n_0}(t)=\la N,n_0|\psi(t)\ra=\la N,n_0|e^{-iH_ft}|\varepsilon_0\ra
  =\sum_k e^{-iE_k^ft}\la N,n_0|\varepsilon^f_k\ra\la\varepsilon^f_k|\varepsilon_0\ra,
\ee
with $|\varepsilon_k^f\ra$ being the $k$th eigenstate of 
$H_f$ and $E_k^f$ is the corresponded energy.
Then, the finite-size fractal dimensions of $|\psi(t)\ra$ is given by
\be \label{FrDt}
  D_q(t)=\frac{S_q(t)}{\ln\mathcal{N}},\quad \text{with}\quad 
  S_q(t)=\frac{\ln\left(\sum_{n_0}|c_{n_0}(t)|^{2q}\right)}{1-q}.
\ee
Evidently, this expression indicates that the dynamics of $D_q$ are determined by $|c_{n_0}(t)|^2$. 

In Figs.~\ref{CfDqs}(a)-\ref{CfDqs}(c), we plot $|c_{n_0}(t)|^2$ as a function of $t$ and $n_0/N$ for 
the quenches that are below, at, and above the critical value with $\xi_i=1$ and system size $N=1000$.
The difference in the time evolution of $|c_{n_0}(t)|^2$ between the cases with $\delta\xi<\delta\xi_c$
and $\delta\xi>\delta\xi_c$ is clearly visible. 
Specifically, $|c_{n_0}(t)|^2$ exhibits quite regular behavior with slowly spreading over
the Fock states for $\delta\xi<\delta\xi_c$, as seen in Fig.~\ref{CfDqs}(a).
However, for $\delta\xi>\delta\xi_c$, 
although $|c_{n_0}(t)|^2$ initially shows regular oscillation, it quickly evolves into  
a complex distribution over a quite large number of Fock states, 
as illustrated in Fig.~\ref{CfDqs}(c).     
At $\delta\xi=\delta\xi_c$, as demonstrated in Fig.~\ref{CfDqs}(b), 
$|c_{n_0}(t)|^2$ soon spreads asymmetrically in an obviously irregular fashion, 
which results in quite complicated evolution behavior.
From the result in Fig.~\ref{CfDqs}(b), we also note that the 
largest component of $|c_{n_0}(t)|^2$ locates around the Fock 
state $|N,N\ra$ during the late-time evolution.
The particular behavior of $|c_{n_0}(t)|^2$ exhibited at the critical point 
is distinct from the cases with $\delta\xi\neq\delta\xi_c$.
It also highlights the remarkable impact of ESQPT on the structure of time evolved state.
These observed features of $|c_{n_0}(t)|^2$ lead us to expect that the 
fractal dimensions $D_q\ (q=1,2,\infty)$ of the time evolved state 
should bear a notable change when we straddled the critical point of ESQPT.

  \begin{figure}
  \includegraphics[width=\columnwidth]{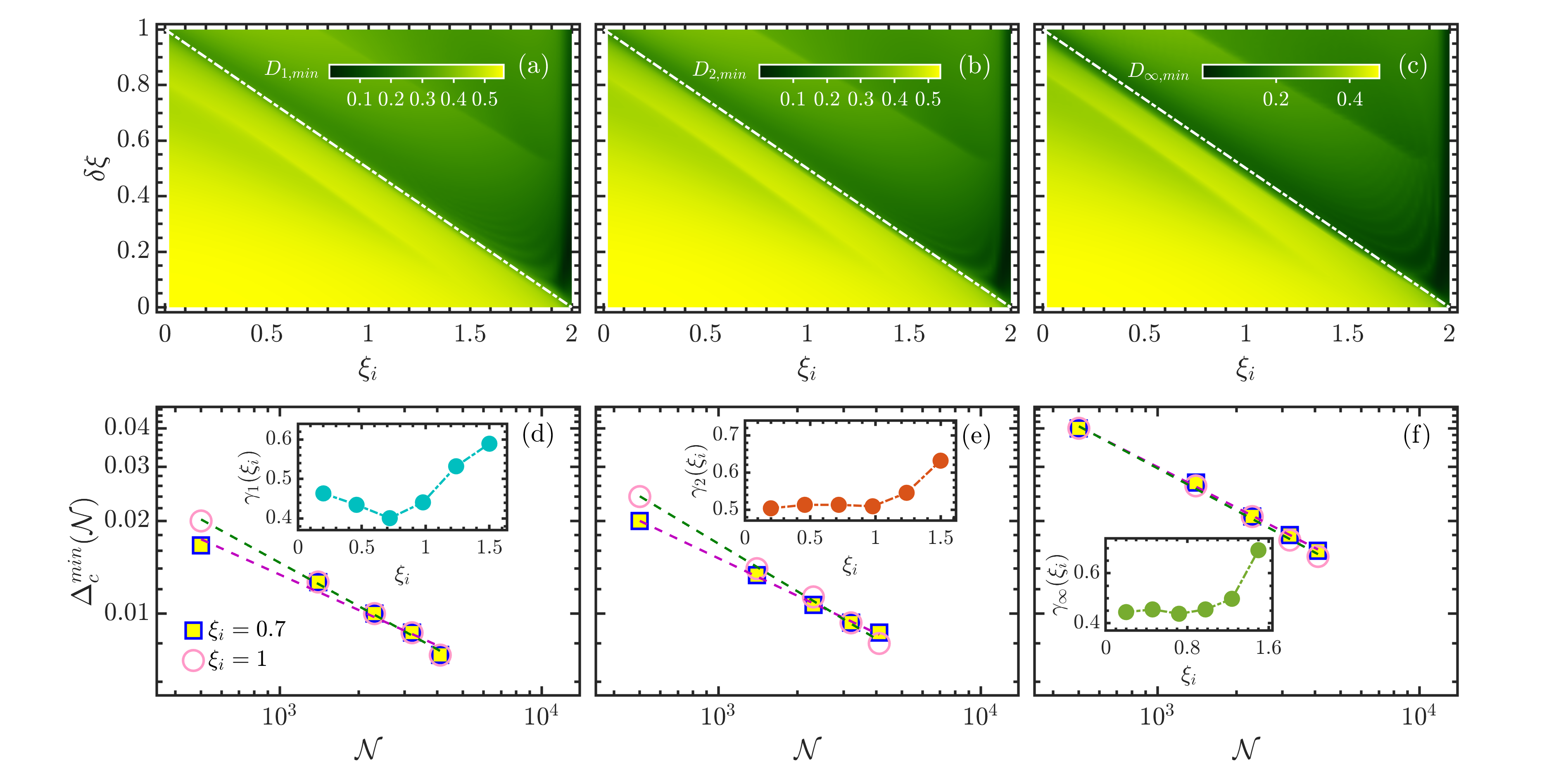}
  \caption{(a)-(c): $D_{q,min}$ in Eq.~(\ref{minimalDq}) as a function of $\xi_i$ and $\delta\xi$
  for $q=1$ (a), $q=2$ (b) and $q=\infty$ (c) with system size $N=1000$. 
  The white dot-dashed line in panels (a)-(c) marks the critical line of ESQPT [see Eq.~(\ref{Cquench})].
  (d)-(f): Scaling of $\Delta_c^{min}(\mathcal{N})=|\delta\xi_c^{min}(\mathcal{N})-\delta\xi_c|$
  with Hilbert space dimension $\mathcal{N}$ for different values of $\xi_i$ 
  [see the legend in panel (d)] and $q=1$ (d), $q=2$ (e), and $q=\infty$ (f).
  Here, $\delta\xi_c^{min}(\mathcal{N})$ is the finite-size precursor of ESQPT critical point 
  and defined by the position of the minimal value of $\partial D_{q,min}/\partial (\delta\xi)$.
  The dashed lines in panels (d)-(f) denote power-law scaling
  $\Delta_c^{min}\sim \mathcal{N}^{-\gamma_q(\xi_i)}$.
  The variation of the scaling exponents $\gamma_q(\xi_i)$ with increasing $\xi_i$
  for $q=1,2$ and $\infty$ are, respectively, plotted in the insets of panels (d)-(f).
  }
  \label{MinDq}
 \end{figure}

The time dependence of $D_q(t)\ (q=1,2,\infty)$ in Eq.~(\ref{FrDt}) for 
the different quench strengths
with $\xi_i=1$ and $N=1000$ are plotted in Figs.~\ref{CfDqs}(d)-\ref{CfDqs}(f).
Several observations are in order.
First, the overall dynamical behaviors of $D_q(t)$ are similar.
Second, $D_q(t)$ increase with time, with smaller oscillations, until it
periodically oscillates around a certain value for $\delta\xi<\delta\xi_c$ case. 
This can be recognized as a consequence of the regular 
and slowly spreading motion in $|c_{n_0}(t)|^2$, as shown Fig.~\ref{CfDqs}(a).
Above the critical point ($\delta\xi>\delta\xi_c$), $D_q(t)$ also increases with time, but with
larger oscillations due to the localized behavior in 
the initial evolution of $|c_{n_0}(t)|^2$, as evidenced in Fig.~\ref{CfDqs}(c).
Moreover, as $|c_{n_0}(t)|^2$ evolves into a nonuniform 
distribution over a large number of Fock states, 
$D_q(t)$ finally saturate at a static value with $q$-dependent fluctuations.  
Third, for critical quench with $\delta\xi=\delta\xi_c$, after an initial sudden decrease,
$D_q(t)$ show a fast growth which rapidly approaches to a saturation value 
with erratic fluctuations. 
The initial dip in $D_q(t)$ can be attributed to the initial 
localization in the evolution of $|c_{n_0}(t)|^2$ [see Fig.~\ref{CfDqs}(b)]. 
While the fast growth of $D_q(t)$ is a manifestation of the 
rapid spreading motion of $|c_{n_0}(t)|^2$.

These properties of $D_q(t)$ verify the strong affects of the underlying ESQPT 
on the structure of time evolved state. 
They also provide a promising way to dynamically control 
the degree of localization of quantum states via ESQPTs.
The dynamics of $D_q(t)$ not only helps us to reveal the impacts of ESQPT, but also 
allows us to characterize the dynamical signatures of ESQPT. 
We therefore proceed to explore how ESQPT gets reflected in the time evolution of fractal dimensions.
To this end, we perform a detailed examination of the 
short- and long-time behaviors of $D_q(t)$, respectively.
 
As shown in Fig.~\ref{CfDqs}, a prominent feature
 in the short-time evolution of $D_q(t)$ is an abrupt decrease around the critical point of ESQPT.
 This leads us to consider the minimum value of $D_q(t)$, denoted by
 $D_{q,min}$, in an initial time interval $[0,\tau]$,
 \be\label{minimalDq}
    D_{q,min}=\mathop{\min}_{t\in[0,\tau]}[D_q(t)],
 \ee 
 where $\tau$ should be larger than the initial timescale.
 In our numerical simulation, we take $\tau=6$.
 We have carefully checked that further increasing $\tau$ 
 does not substantially change our main results.  
 
  \begin{figure}
  \includegraphics[width=\columnwidth]{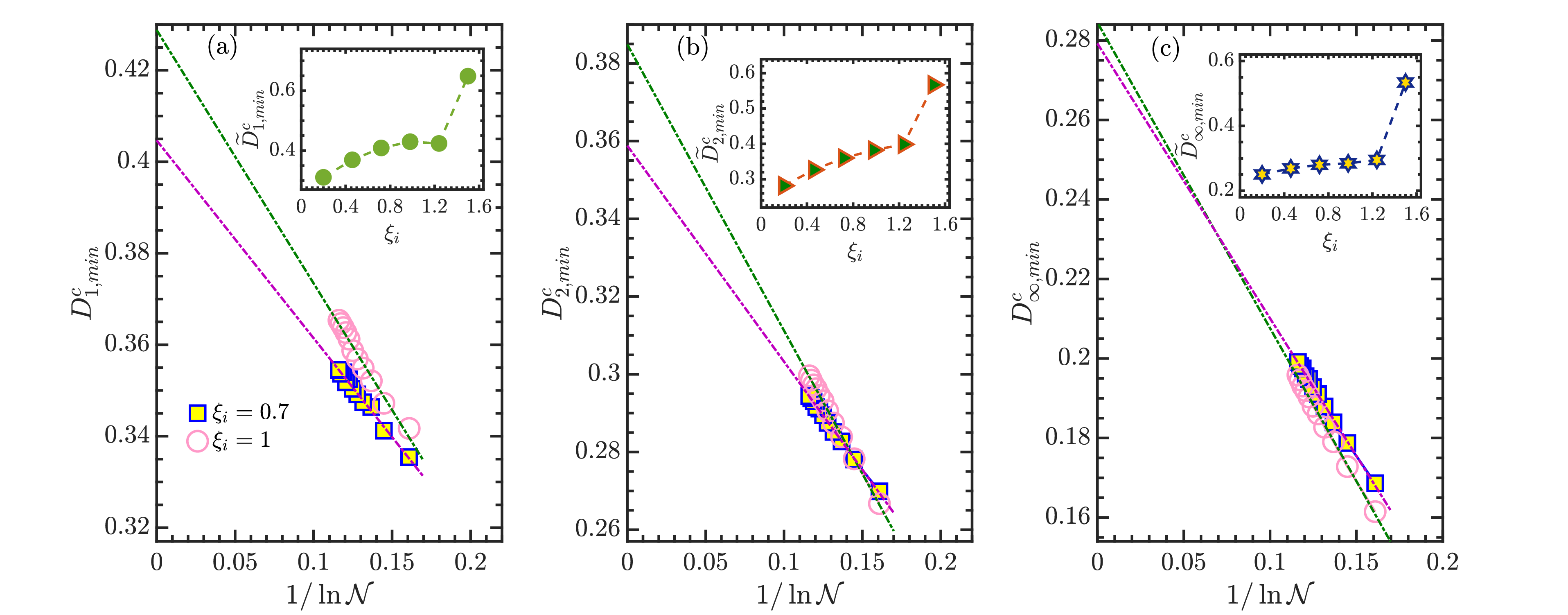}
  \caption{(a)-(c): Finite-size fractal dimensions at the critical value $\delta\xi_c$, 
  $D_{q,min}^c=D_{q,min}(\delta\xi_c)$ as a function of 
  $1/\ln\mathcal{N}$ for different values of $\xi_i$ [see the legend in panel (a)] 
  with $q=1$ (a), $q=2$ (b), and $q=\infty$ (c). 
  Here, $\mathcal{N}$ is the dimension of the Hilbert space.
  The dot-dashed lines in each panel represent the linear function of the form
  $D_{q,min}^c=r_q(\xi_i)/\ln\mathcal{N}+\widetilde{D}_{q,min}^c(\xi_i)$.
  The dependence of critical fractal dimensions $\widetilde{D}_{q,min}^c(\xi_i)$ with
  $\xi_i$ are shown in the insets of panels (a)-(c).
  }
  \label{MinDvsN}
 \end{figure}
 
 In Figs.~\ref{MinDq}(a)-\ref{MinDq}(c), we plot how $D_{q,min}$ varies as a function of 
 $\xi_i$ and $\delta\xi$ for $q=1, 2$, and $\infty$, respectively.
 We see that $D_{q,min}$ undergoes a remarkable decrease around 
 ESQPT critical point, regardless of the value of $q$. 
 This means that the presence of ESQPT can be detected by $D_{q,min}$.
 To further confirm this statement, we define the finite-size precursor 
 of ESQPT critical point, $\delta\xi_c^{min}(\mathcal{N})$,
 as the minimal value of $\partial D_{q,min}/\partial(\delta\xi)$ and analyze
 how it converges to the exact value $\delta\xi_c$ in Eq.~(\ref{Cquench}) by considering
 $\Delta_c^{min}=|\delta\xi_c^{min}(\mathcal{N})-\delta\xi_c|$ as a function of $\mathcal{N}$.
For different values of $q$, the scaling of $\Delta_c^{min}$ with 
Hilbert space dimension $\mathcal{N}$ for two representative values of $\xi_i$ are shown in
Figs.~\ref{MinDq}(d)-\ref{MinDq}(f).
Obviously, $\Delta_c^{min}$ exhibits a power-law scaling 
$\Delta_c^{min}\sim\mathcal{N}^{-\gamma_q(\xi_i)}$, which is independent of the $q$ value.
However, the dependence of the exponent $\gamma_q(\xi_i)$ on $q$ and $\xi_i$ is clearly visible. 
In the insets of Figs.~\ref{MinDq}(d)-\ref{MinDq}(f), we demonstrate
how the scaling exponent $\gamma_q(\xi_i)$ evolves as a function of $\xi_i$ for 
the corresponding values of $q$.
Moreover, we also reveal that the fractal dimensions $D_{q,min}$ at ESQPT critical point, 
denoted by $D_{q,min}^c$, behaves as a linear function of $1/\ln\mathcal{N}$,
$D_{q,min}^c=r_q(\xi_i)/\ln\mathcal{N}+\widetilde{D}^c_{q,min}(\xi_i)$, 
as demonstrated in Figs.~\ref{MinDvsN}(a)-\ref{MinDvsN}(c).
This results in the $\xi_i$-dependent critical fractal dimensions $\widetilde{D}_{q,min}^c (\xi_i)$ 
in the thermodynamic limit $\mathcal{N}\to\infty$.
The variation of $\widetilde{D}_{q,min}^c(\xi_i)$ 
with increasing $\xi_i$ for different values of $q$ are 
shown in the insets of Figs.~\ref{MinDvsN}(a)-\ref{MinDvsN}(c).

To characterize ESQPT through the long-time properties of $D_q(t)$, we consider
the probability distribution for the scaled fractal dimensions 
in a certain time window $[t_s,t_s+T]$,
\be \label{Pdfsfd}
  P_q(\mathcal{D})=\lim_{T\to\infty}\frac{1}{T}\int_{t_s}^{t_s+T}
     \delta[\mathcal{D}-\mathcal{D}_q(t)]dt,
\ee 
where $t_s$ should be sufficient larger than the initial timescale.
Here, the scaled finite-size fractal dimensions $\mathcal{D}_q(t)$ are defined as
\be
   \mathcal{D}_q(t)=\frac{D_q(t)-D_q^{min}}{D_q^{max}-D_q^{min}},
\ee
with $D_q^{min}=\mathop{\min}_{t\in[t_s,t_s+T]}[D_q(t)]$ and
$D_q^{max}=\mathop{\max}_{t\in[t_s,t_s+T]}[D_q(t)]$.
The definition of $\mathcal{D}_q(t)$ means $\mathcal{D}_q(t)\in[0,1]$.  
In our numerical calculation, we take $t_s=100$ and $T=1000$. 
A careful check shows that our main results are still hold for further increasing $t_s$ and $T$.  
The cumulative distribution of $\mathcal{D}$ is calculated by
\be\label{CdfFd}
   F_q(\mathcal{D})=\int_0^{\mathcal{D}}P_q(x)dx,
\ee
with $P_q(x)$ being the probability distribution of 
the scaled finite-size fractal dimensions in Eq.~(\ref{Pdfsfd}).

 \begin{figure}
  \includegraphics[width=\columnwidth]{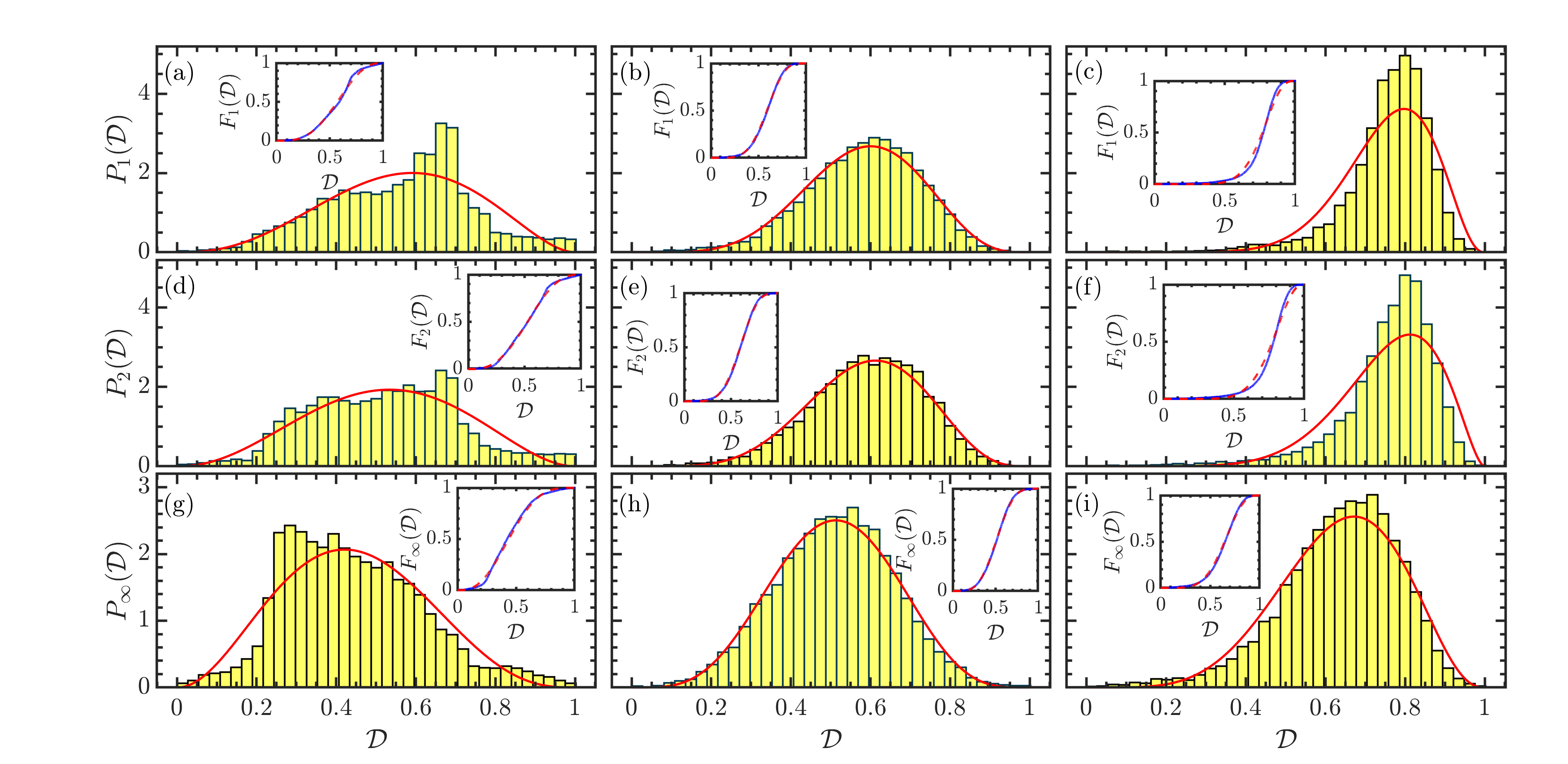}
  \caption{(a)-(c): Probability distribution of the scaled $\mathcal{D}_1(t)$,
  $P_1(\mathcal{D})$, for $\delta\xi=0.1$ (a),
  $\delta\xi=0.5$ (b), and $\delta\xi=1$ (c).
  (d)-(f): Probability distribution of $\mathcal{D}_2(t)$,
  $P_2(\mathcal{D})$, for the same values of $\delta\xi$ as in panels (a)-(c).
  (g)-(i): $P_\infty(\mathcal{D})$ for different $\delta\xi$ values that are the 
  same as those to panels (a)-(c).   
  In all cases, the control parameter is fixed to $\xi_i=1$, which 
  gives rise to $\delta\xi_c=0.5$ [cf.~Eq.~(\ref{Cquench})], while the system size is $N=1000$.
  The red solid curve in each panel denotes the optimized beta distribution in Eq.~(\ref{betaPdf}) 
  with shape parameters $(a,b)$ are (a) $(3.7140, 2.8764)$, (b) $(6.5863, 4.7199)$, (c) $(10.7023, 3.4622)$, 
  (d) $(3.2958, 3.0046)$, (e) $(6.5683, 4.5281)$, (f) $(8.7197, 2.7828)$, 
  (g) $(3.1217, 3.9167)$, (h) $(5,2670, 5.0571)$, and (i) $(6.1505, 3.5182)$.
  The inset in each panel compares the cumulative distribution function of scaled fractal dimensions
  $F_q(\mathcal{D})$ (blue solid line) to the corresponding result 
  of the beta distribution (red dashed line). 
  }
  \label{Prbdist}
 \end{figure}

The computed probability distribution of $\mathcal{D}_q(t)$ and 
the associated cumulative distribution for $q=1, 2$, and $\infty$ are shown
in Fig.~\ref{Prbdist}, with $\xi_i=1$ and $N=1000$.  
In all cases, we plot the values of $\delta\xi$ below, at, 
and above the critical value $\delta\xi_c$.
For the values of $\delta\xi$ that are far away from the critical value, 
irrespective of the values of $q$, the distribution of $P_q(\mathcal{D})$ 
is an asymmetry distribution with heavy-tail, in particular for the case of $\delta\xi>\delta\xi_c$.
This stems from the asymmetrical fluctuations around a saturation value 
exhibited by the long-time evolution of $D_q(t)$. 
However, once $\delta\xi=\delta\xi_c$, the shape of $P_q(\mathcal{D})$ becomes more symmetry with light-tild.
This reflects the fact that the fluctuations in long-time evolution of $D_q(t)$ 
at $\delta\xi=\delta\xi_c$ are more symmetrical than the cases of $\delta\xi\neq\delta\xi_c$. 

The fact of $\mathcal{D}_q(t)\in[0,1]$ inspires us to fit $P_q(\mathcal{D})$ by a beta
distribution \cite{Gupta2004},
\be \label{betaPdf}
   f_B(x)=\frac{x^{a-1}(1-x)^{b-1}}{B(a,b)},
\ee
where $B(a,b)=\int_0^1v^{a-1}(1-v)^{b-1}dv$ is the beta 
function and $a, b>0$ are the shape parameters.
The associated cumulative distribution function of $f_B(x)$ is given by
\be
   F_B(x)=\int_0^x f(y)dy=I_x(a,b),
\ee
where $I_x(a,b)=B(x;a,b)/B(a,b)$ is the regularized incomplete beta function
with $B(x;a,b)=\int_0^x y^{a-1}(1-y)^{b-1}dy$ being the incomplete beta function. 

In Fig.~\ref{Prbdist}, the best fitted beta distribution in Eq.~(\ref{betaPdf}) 
and the corresponding cumulative distribution for each case are respectively plotted
as the red solid and dashed curves in the main panel and inset. 
The obvious deviation between $P_q(\mathcal{D})$ and $f_B(x)$ can be immediately
identified for $\delta\xi\neq\delta\xi_c$ cases, 
as shown in the first and last columns of Fig.~\ref{Prbdist}. 
However, the distribution $P_q(\mathcal{D})$ is well captured by the beta distribution
at the critical point with $\delta\xi=\delta\xi_c$, 
as seen in the middle column of Fig.~\ref{Prbdist}.
The degree of deviation between $P_q(\mathcal{D})$ and the beta distribution can be quantitatively
examined by the root-mean-square-error (RMSE), which measures the
difference between observed and predicted values \cite{Schervish2014}.
For our studies, the RMSE for the distribution of $\mathcal{D}_q(t)$ is defined as
\be \label{Rmse}
   R_q=\sqrt{\int_0^1\left[F_q(x)-F_B(x)\right]^2dx},
\ee
with $F_q(x)$ being the cumulative distribution of $P_q(\mathcal{D})$ in Eq.~(\ref{CdfFd}).
It vanishes when $P_q(\mathcal{D})=f_B(x)$ and a strong deviation 
of $P_q(\mathcal{D})$ from $f_B(x)$ is indicated by larger RMSE.

 \begin{figure}
  \includegraphics[width=\columnwidth]{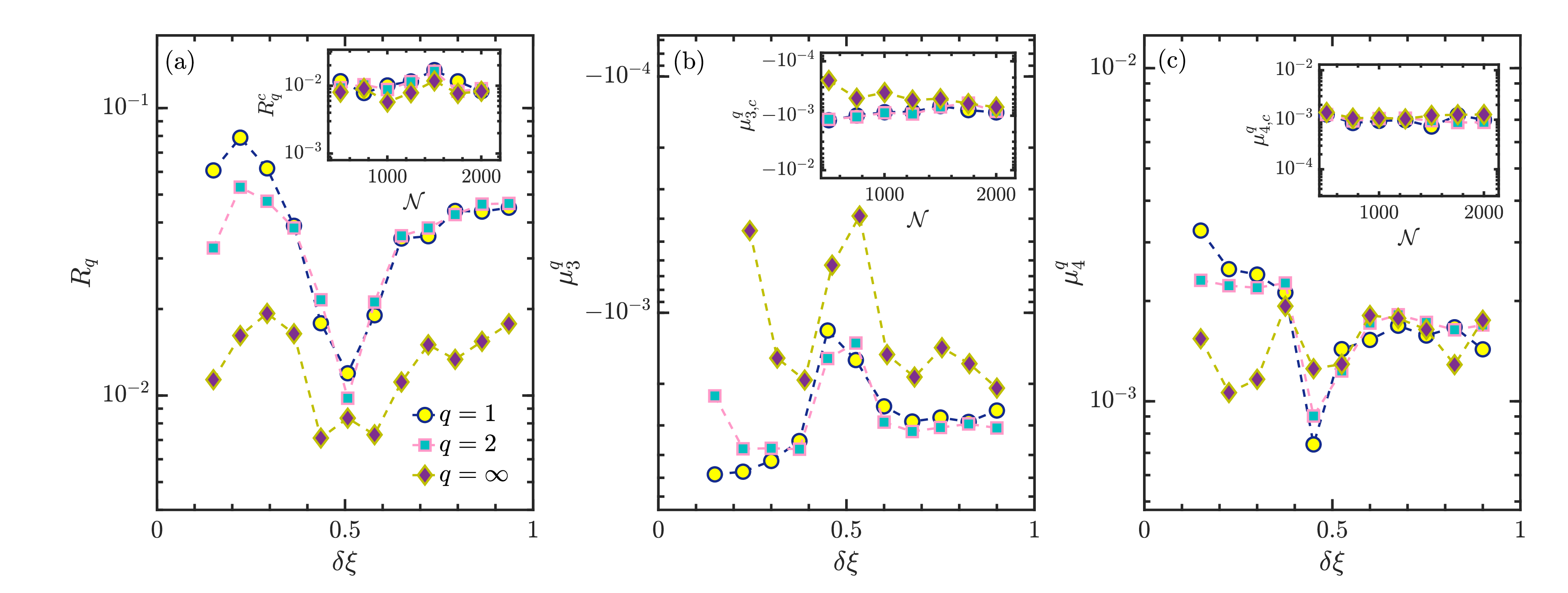}
  \caption{(a): $R_q$ in Eq.~(\ref{Rmse}) as a function of $\delta\xi$ for different values of $q$.
  The inset shows how the critical value of RMSE, defined as $R_q^c=R_q(\delta\xi_c)$, varies
  as a function of the Hilbert space dimension $\mathcal{N}$ for different $q$ values.
  (b)-(c): Third (b) and fourth (c) central moments of $P_q(\mathcal{D})$ as a function of
  $\delta\xi$ for the same values of $q$ as in panel (a).
  Insets in panels (b) and (c) show $\mu_3^q$ and $\mu_4^q$ as a function of $\mathcal{N}$
  for $\delta\xi=\delta\xi_c$ with $q=1,2$ and $\infty$ [see the legend in panel (a)].
 Other parameters: $N=1000$, $\xi_i=1$, and $\delta\xi_c=0.5$, obtained from Eq.~(\ref{Cquench}). 
  }
  \label{ScalingMus}
 \end{figure}

 The dependence of $R_q\ (q=1,2,\infty)$ with $\delta\xi$ for
 system size $N=1000$ and $\xi_i=1$ are plotted in Fig.~\ref{ScalingMus}(a).
 Regardless of the value of $q$, an obvious dip near the critical value $\delta\xi_c$
 in the behaviors of $R_q$ can be clearly identified. 
 This implies that the distribution $P_q(\mathcal{D})$ is well described by the beta distribution
 around the critical point of ESQPT, as already observed in Fig.~\ref{Prbdist}. 
 Moreover, at the critical point, we find that $R_q$ evolves 
 around a vanishingly small value with a tiny fluctuation as the system size is increased,
 as seen in the inset of Fig.~\ref{ScalingMus}(a).   
 
 Further signatures of ESQPT can be revealed by the statistical properties of $P_q(\mathcal{D})$.
 We therefore investigate the $n$th central moment of $P_q(\mathcal{D})$,
 \be
    \mu_n^q=\int_0^1P_q(\mathcal{D})(\mathcal{D}-\overline{\mathcal{D}})^nd\mathcal{D},
\ee
where $\overline{\mathcal{D}}=\int_0^1\mathcal{D}P_q(\mathcal{D})d\mathcal{D}$ is the
averaged $\mathcal{D}$.
In this work, we focus on the central moments with $n=3$ and $4$, known as the
skewness and kurtosis of the distribution. 

It is known that the third and fourth central moments measure the
lopsidedness and the heaviness of the tail of a distribution.
The negative third central moment corresponding to the distributions that are skewed to the left,
while it takes positive values for the right skew distributions. 
In particular, the symmetric distributions are indicated by zero third central moment. 
The fourth central moment is always positive as it defines as an expectation of the fourth power. 
A light-tailed distribution is usually associated with a smaller fourth central moment. 
The observed features of $P_q(\mathcal{D})$ in Fig.~\ref{Prbdist} allow us to expect that
both $\mu_3^q$ and $\mu_4^q$ would exhibit a particular behavior at the critical point of ESQPT.
This is verified in Figs.~\ref{ScalingMus}(b) and \ref{ScalingMus}(c),
where the variations of the central moments with $\delta\xi$ are, respectively, plotted
for $N=1000$ and $\xi_i=1$.
It is evident that $\mu_3^q$ and $\mu_4^q$ show local peaks and sudden dips 
in the neighborhood of the critical point of ESQPT, regardless of the value of $q$.
This indicates that $P_q(\mathcal{D})$ close to a symmetric 
and light-tailed distribution around the critical value $\delta\xi_c$,
in according with the results demonstrated in the middle column of Fig.~\ref{Prbdist}.
We also find that the values of $\mu_3^q$ and $\mu_4^q$ at the 
critical value $\delta\xi_c$ are almost independent of the system size, 
as shown in the insets of Figs.~\ref{ScalingMus}(b) and \ref{ScalingMus}(c).

\begin{figure}
  \includegraphics[width=\columnwidth]{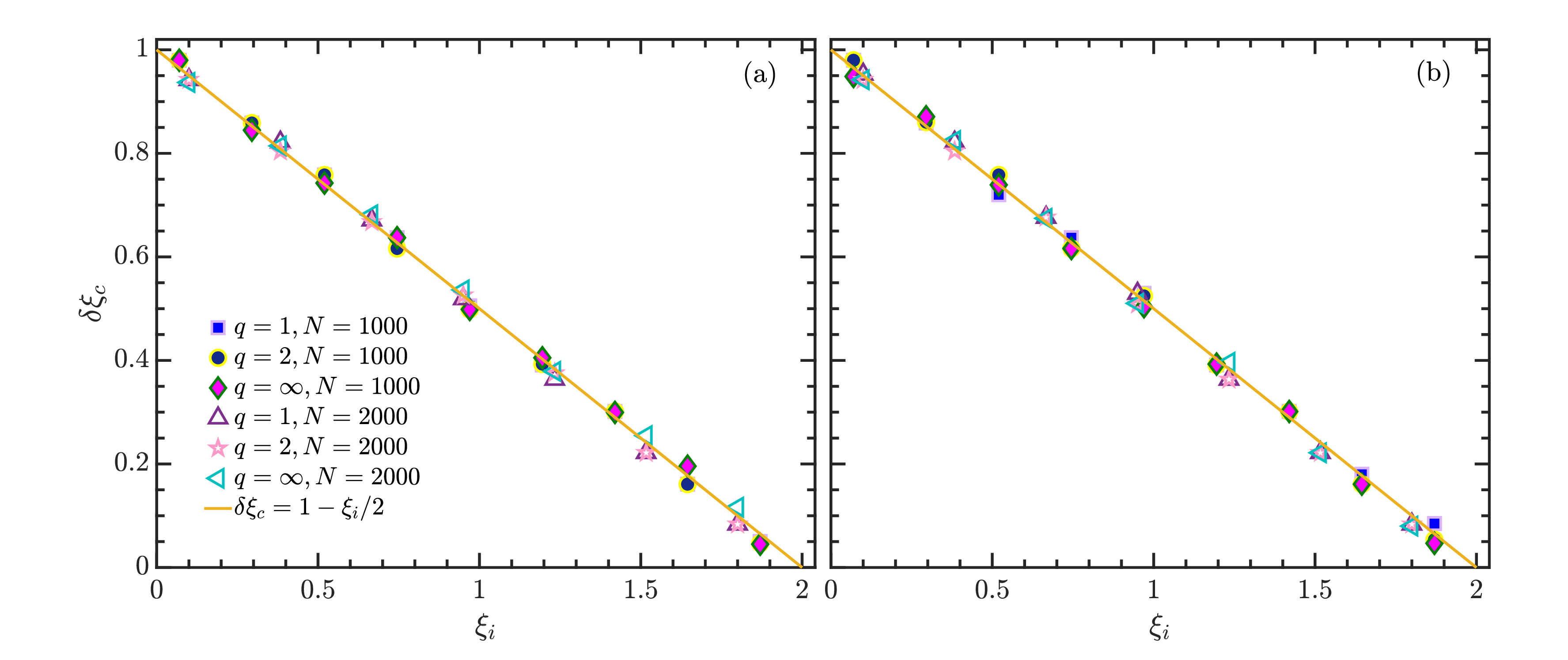}
  \caption{(a)-(b): Estimated critical value $\delta\xi_c$, extracted from $\mu_3^q$ (a) and
  $\mu_4^q$ (b) as a function of $\xi_i$ for different values of $q$ and system sizes $N$
  [see the legend in panel (a)].
  For both $\mu_3^q$ and $\mu_4^q$, the critical value $\delta\xi_c$ is 
  estimated by the position of their local extreme value. 
  The solid line represents the analytical result in Eq.~(\ref{Cquench}). 
  }
  \label{EstimationCrt}
 \end{figure}

The local peaks and dips exhibited by the central moments 
suggest that their positions can be employed as an estimation of the 
critical point of ESQPT for finite-size systems. 
The estimated critical value $\delta\xi_c$ obtained from $\mu_3^q$ and $\mu_4^q$ 
as a function of $\xi_i$ for different values of $q$ and system sizes 
are plotted in Figs.~\ref{EstimationCrt}(a) and \ref{EstimationCrt}(b), respectively.
For each case, the analytical result of $\delta\xi_c$ in Eq.~(\ref{Cquench}) 
is also plotted as the solid line. 
We see a good agreement between the numerical and analytical results and the 
agreement is enhanced by increasing the system size.
This confirms that the presence of ESQPT can be reliably probed by 
the central moments of $P_q(\mathcal{D})$.

We finally explore what are the affects and signatures of ESQPT 
in the structure of the long-time averaged state, which is defined as
\be
  |\Psi\ra=\lim_{T\to\infty}\frac{1}{T}\int_0^Tdt |\psi(t)\ra
     =\lim_{T\to\infty}\frac{1}{T}\int_0^Tdt \sum_kd_ke^{-iE_k^ft}|\varepsilon_k^f\ra,
\ee
with $d_k=\la\varepsilon_k^f|\varepsilon_0\ra$ being the overlap 
between the $k$th eigenstate of $H_f$ and the initial state. 
In the Fock basis, $|\Psi\ra$ can be expanded as
\be
  |\Psi\ra=\sum_{n_0}\mathcal{X}_{n_0}|N,n_0\ra,
\ee 
where the expansion coefficients are given by
\be
  \mathcal{X}_{n_0}=\la N,n_0|\Psi\ra
   =\lim_{T\to\infty}\frac{1}{T}\int_0^T dt\sum_kd_ke^{-iE_k^ft}\la N,n_0|\varepsilon_k^f\ra.
\ee
As the spectrum of $H$ (\ref{SpinH}) has no degenerate, it is straightforward to find that
\be
  |\mathcal{X}_{n_0}|^2=\sum_k|d_k|^2|\la N,n_0|\varepsilon_k^f\ra|^2.
\ee
The finite-size fractal dimensions of $|\Psi\ra$ are calculated as
\be
   \overline{D}_q=\frac{\overline{S}_q}{\ln\mathcal{N}},\quad \text{with}\quad
   \overline{S}_q=\frac{\ln\left(\sum_{n_0}|\mathcal{X}_{n_0}|^{2q}\right)}{1-q}.
\ee  

 \begin{figure}
  \includegraphics[width=\columnwidth]{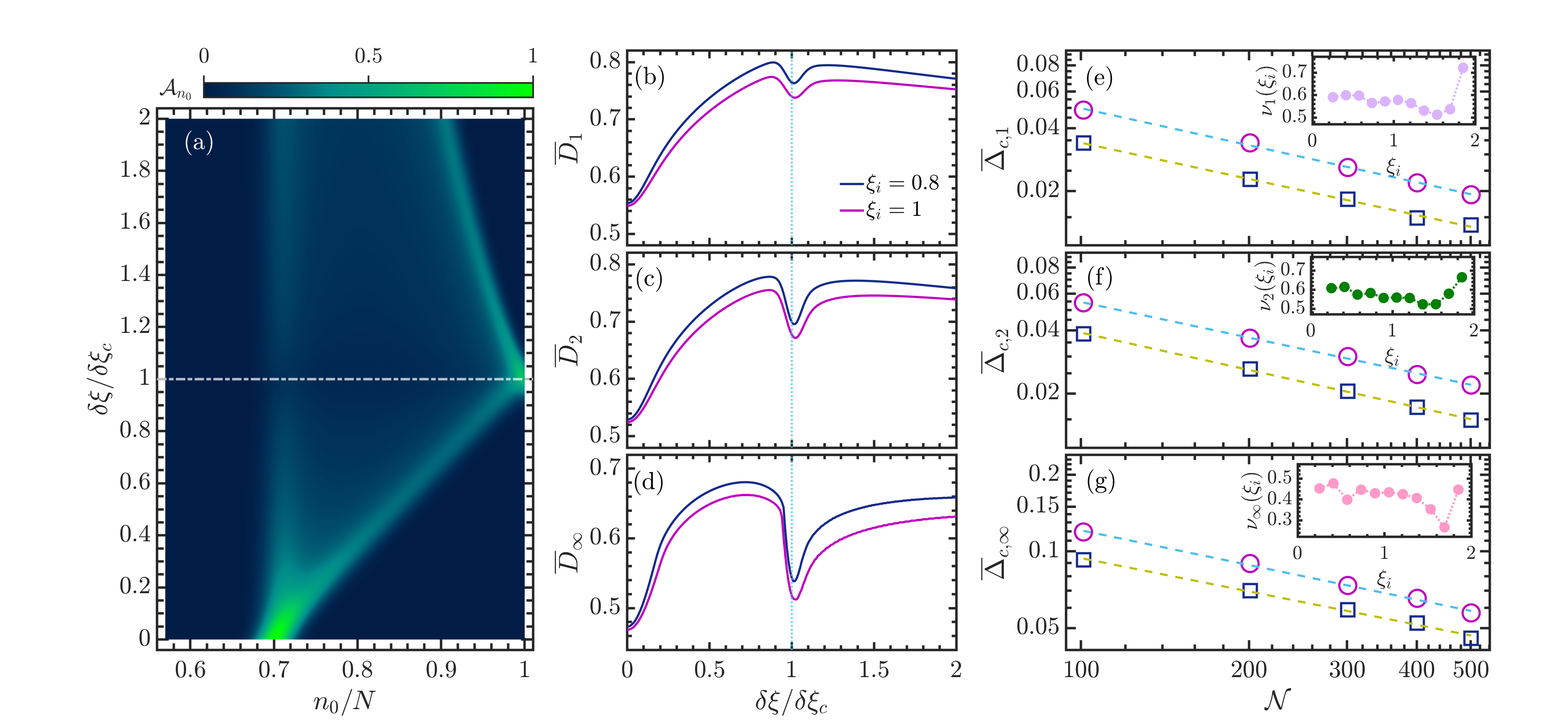}
  \caption{(a) Rescaled expansion coefficients, 
  $\mathcal{A}_{n_0}=|\mathcal{X}_{n_0}|^2/\mathcal{X}_{max}$, as a function
  of $\delta\xi/\delta\xi_c$ and $n_0/N$ with system size $N=1000$. 
  Here, $\mathcal{X}_{max}=\mathrm{Max}(|\mathcal{X}_{n_0}|^2)$ 
  is the maximal value of $|\mathcal{X}_{n_0}|^2$.
  The white horizontal dot-dashed line marks the critical value $\delta\xi_c$.
   (b)-(c): Finite-size fractal dimensions $\overline{D}_1$ (a), $\overline{D}_2$ (b), and
   $\overline{D}_\infty$ (c) as a function of $\delta\xi/\delta\xi_c$ for
   different values of $\xi_i$ [see the legend in panel (b)] with system size $N=1000$. 
   The vertical dotted line in each panel denotes the critical point of ESQPT.
   (e)-(g): Distance between the estimated and analytical critical points, 
   $\overline{\Delta}_{c,q}=|1-\delta\xi_{dip,q}/\delta\xi_c|$, 
   versus the Hilbert space dimension $\mathcal{N}$ for 
   $\xi_i=0.8$ (square symbols) and $\xi_i=1$ (circle symbols) 
   with $q=1$ (e), $q=2$ (f), and $q=\infty$ (g).
   Here, $\delta\xi_{dip,q}/\delta\xi_c$ denotes the position of dip in $\overline{D}_q$.
   The dashed lines in each panel indicate
   $\overline{\Delta}_{c,q}\sim\mathcal{N}^{-\nu_q(\xi_i)}$.
   The dependence of $\nu_q(\xi_i)$ on the control parameter $\xi_i$ for $q=1,2,\infty$
   are, respectively, plotted in the insets of panels (e)-(g).
   }
  \label{Fdvsxi}
 \end{figure}

In Fig.~\ref{Fdvsxi}(a), we show how the scaled expansion coefficients vary as a function of the scaled
quench strength $\delta\xi/\delta\xi_c$ and $n_0/N$.
We see that $|\Psi\ra$ spreads out in the Fock basis with increasing $\delta\xi$
until the critical value $\delta\xi_c$ is reached.
Above the critical point of ESQPT, the number of Fock states 
occupied by $|\Psi\ra$ is decreased with increasing $\delta\xi$. 
In particular, at $\delta\xi=\delta\xi_c$, $|\Psi\ra$ 
is highly localized in $|N,N\ra$ state.
One can therefore expect that a significant decrease around the critical point of ESQPT 
should be presented in the behavior of $\overline{D}_q$, irrespective of the value of $q$. 
The dependence of $\overline{D}_q$ with $\delta\xi/\delta\xi_i$ for $q=1,2$, and $\infty$ are, respectively,
plotted in Figs.~\ref{Fdvsxi}(b)-\ref{Fdvsxi}(d).
They show that $\overline{D}_q$ exhibit an obvious dip around the critical value $\delta\xi_c$,
indicating that the underlying ESQPT leads to the localization of the long-time averaged state.

\begin{figure}
    \includegraphics[width=\columnwidth]{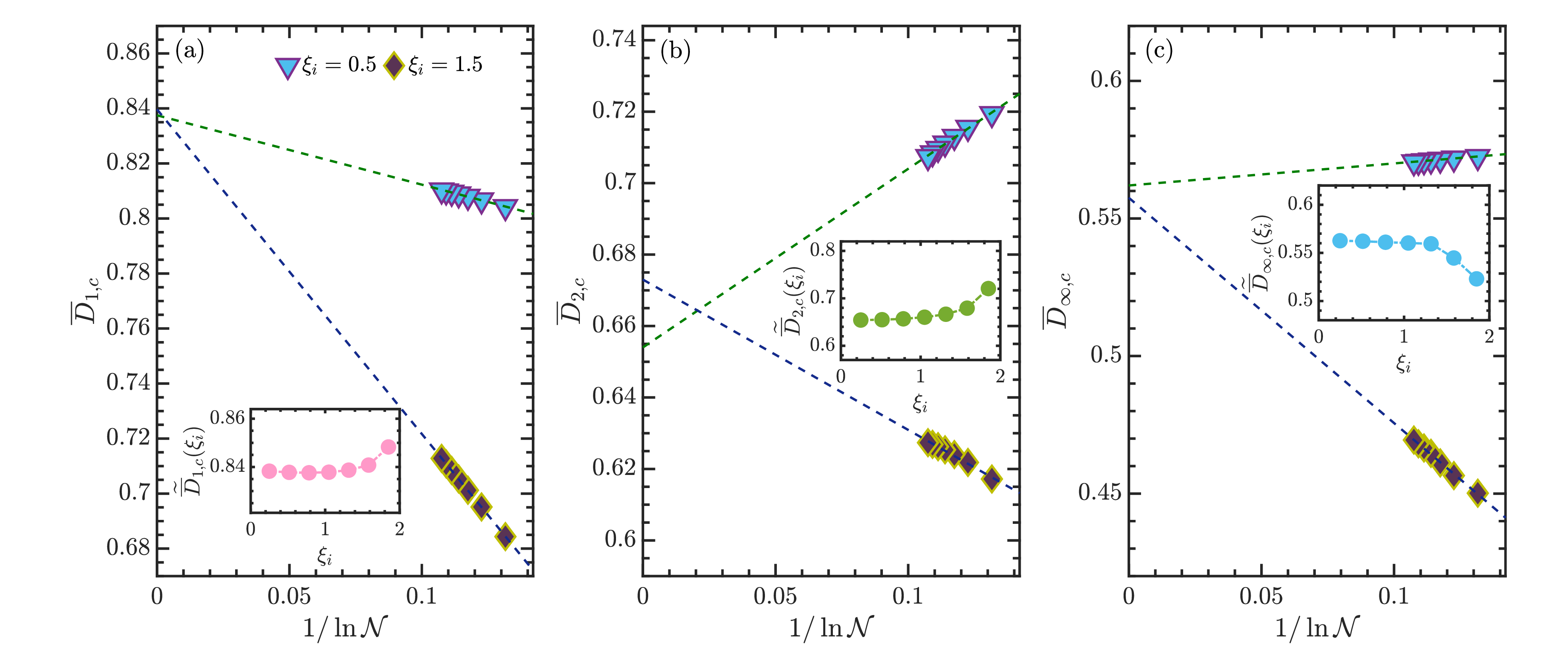}
  \caption{(a)-(c) Scaling of $\overline{D}_{1,c}$ (a), $\overline{D}_{2,c}$ (b), 
  and $\overline{D}_{\infty,c}$ with $1/\ln\mathcal{N}$ 
  for different values of $\xi_i$ [see the legend in panel (a)]. 
  The dashed lines in each panel denote the fitted linear function of the form
  $\overline{D}_{q,c}=a_q(\xi_i)/\ln\mathcal{N}+\widetilde{\overline{D}}_q,c(\xi_i)$.
  The dependence of the fractal dimension $\widetilde{\overline{D}}_{q,c}(\xi_i)$ with $\xi_i$
  for $q=1,2$ and $\infty$, are shown in the insets of main panels.
   }
  \label{ScalingFDs}
 \end{figure}

The observed dips in $\overline{D}_q$ with $q=1,2,\infty$ not only reveal the impacts 
of ESQPT on the state $|\Psi\ra$, but also enable us to detect the presence of ESQPT.
To confirm this, we take the position of the dip, denoted by $\delta\xi_{dip,q}/\delta\xi_c$, 
as an estimation of the critical point of ESQPT, and
study how it tends to the exact one as the system size is increased.
Thus, we investigate the scaling of the distance between the estimated and exact critical points,
$\overline{\Delta}_{c,q}=|1-\delta\xi_{dip,q}/\delta\xi_c|$, 
with the Hilbert space dimension $\mathcal{N}$.
As shown in Figs.~\ref{Fdvsxi}(e)-\ref{Fdvsxi}(g), 
$\overline{\Delta}_{c,q}$ scales as
$\overline{\Delta}_{c,q}\sim\mathcal{N}^{-\nu_q(\xi_i)}$, 
independent of the value of $q$.
However, the critical point $\nu_q(\xi_i)$
shows a strong dependences on $q$ and $\xi_i$, as seen
in the insets of Figs.~\ref{Fdvsxi}(e)-\ref{Fdvsxi}(g), where we plot the 
variation of $\nu_q(\xi_i)$ with $\xi_i$ for $q=1,2$, and $\infty$, respectively.
These results allow us to claim that the fractal dimensions of the long-time averaged state
serve as the powerful probes of ESQPT.
In Fig.~\ref{ScalingFDs}, we demonstrate how the value of $\overline{D}_q$ 
at $\delta\xi_c$, denoted by $\overline{D}_{q,c}$, varies as a function of $1/\ln\mathcal{N}$
for different values of $\xi_i$ and $q$.
It can be clearly seen that the variation of $\overline{D}_{q,c} (q=1,2,\infty)$ 
with $1/\ln\mathcal{N}$ are well captured by a linear function of the form 
$\overline{D}_{q,c}=a_q(\xi_i)/\ln\mathcal{N}+\widetilde{\overline{D}}_{q,c}(\xi_i)$, 
regardless of the value of $\xi_i$.
Here, $\widetilde{\overline{D}}_{q,c}(\xi_i)=\lim_{\mathcal{N\to\infty}}\overline{D}_{q,c}$ 
are the the fractal dimensions and depend on the control parameter $\xi_i$, as
observed in the insets of Fig.~\ref{ScalingFDs}, where we plot 
$\widetilde{\overline{D}}_{q,c}$ as a function of $\xi_i$ for each value of $q$.
This provides further evidence of the usefulness of the fractal dimensions for studying of
ESQPT and also verifies the localization effect of ESQPT.

\section{Conclusions}\label{Fourth}

In summary, we have performed a detail examination of the affects of ESQPT on the multifractality of
quantum states by means of the multifractal analysis. 
The impacts of ESQPT on the multifractality of eigenstates have been 
discussed in several previous works via the inverse participation ratio 
\cite{Santos2015,Santos2016,Gamito2022}.
Here, we have extended those studies to analyze the multifractal dimensions of 
both eigenstates and time evolved state.

Our explorations are carried out in a ferromagnetic spin-$1$ BEC, which is 
a highly tunable platform and has been employed as a prototypical model 
for theoretical \cite{Feldmann2021} and experimental \cite{Meyer2023} studies of ESQPTs.  
We have shown that the multifractality of both eigenstates and time evolved states are strongly affected
by the presence of ESQPT.
For the eigenstates, it has been found a sudden dip in the behavior of 
their finite-size fractal dimensions, indicating a strong localization effect of ESQPT.
Further scaling analysis have revealed that the fractal dimensions of the critical eigenstate vary
in the interval $0<\widetilde{D}_{q,c}<1$  for $q>0$ and decreases with increasing $q$. 
We have also demonstrated that the eigenstate localization induced by ESQPT 
can also be used as an efficient detector of ESQPT.

Regarding the time evolved state, we have shown that the affects of ESQPT 
on its multifractality are manifested by different time evolution 
behaviors of the finite-size fractal dimensions.
Moreover, the particular behavior of the fractal dimensions 
at the critical point not only highlights the impacts, but also
indicates the occurrence of ESQPT. 
By focusing on the short- and long-time evolution properties of the finite-size fractal dimensions,
we have quantitatively analyzed how ESQPT gets reflected in the 
fractal dimensions of the time evolved state.
We have found that the underlying ESQPT results in an obvious dip 
near the critical point in the short-time behaviors of fractal dimensions.
This fact verifies the localization effect of ESQPT and allows us to 
utilize it as a probe of ESQPT.
Further impacts of ESQPT are unveiled by the probability distribution of the fractal dimensions
defined in a certain long-time interval.  
We have demonstrated that the distribution of fractal dimensions 
undergoes a remarkable change when the system crosses the ESQPT. 
In particular, we have shown that the distribution of fractal dimensions at the critical point of ESQPT
can be well described by the beta distribution. 
This leads us to study the central moments of the distribution of fractal dimensions and 
show that they can reliably capture the occurrence of ESQPT.
We finally discuss how to expose ESQPT through the fractal 
dimensions of long-time averaged state.
Our results unveil that the long-time averaged state is also strongly localized at the critical
point of ESQPT and the observed sudden dip in its fractal dimensions act as valid probes 
to detect the critical point of ESQPT.

A natural extension of the present work is to explore whether our main conclusions are still
hold in other systems with different kinds ESQPTs, such as the Dicke model,
in which the ESQPT is characterized by the singularities in the first derivative of the density of state. 
Another possible direction of research is to analytically explain why 
the distribution of fractal dimensions of the time evolved state follows the
beta distribution at the critical point of ESQPT. 
Our study contributes further understanding of both static and dynamical
features of ESQPT, providing a potential application of ESQPTs in quantum state engineering.
Furthermore, the highly controllable of the spin-$1$ BEC also leads us to hope 
that our findings could motivate more experimental investigations of various aspects of ESQPTs.

 \acknowledgments
  
We acknowledge the support from the Slovenian Research and Innovation Agency (ARIS)
under Grants No.~J1-4387 and No.~P1-0306.
Z.~X.~N acknowledges the financial support from Zhejiang Provinical Nature Science
Foundation under the Grant No.~LQ22A040006.
This work was also support from Zhejiang Provinical Nature Science
Foundation under the Grant No.~LY20A050001.

\bibliographystyle{apsrev4-1}


%

\end{document}